# HRCTCov19 - A High-Resolution Chest CT Scan Image Dataset for COVID-19 Diagnosis and Differentiation

Iraj Abedi[1]
Mahsa Vali[2]
Bentolhoda Otroshi[3]
Maryam Zamanian[1]
Hamidreza Bolhasani[4]*

[1] Department of Medical Physics, School of Medicine, Isfahan University of Medical Sciences, Isfahan, Iran.
[2] Department of Electrical and Computer Engineering, Isfahan University of Technology, Isfahan, Iran.
[3] Department of Radiology, School of Medicine, Arak University of Medical Sciences, Arak, Iran
[4] Department of Computer Engineering, Science and Research Branch, Islamic Azad University, Tehran, Iran.

* Corresponding Author: Hamidreza Bolhasani

Email: hamidreza.bolhasani@srbiau.ac.ir

## Abstract

**Introduction:** During the COVID-19 pandemic, computed tomography (CT) was a popular method for diagnosing COVID-19 patients. HRCT (High-Resolution Computed Tomography) is a form of computed tomography that uses advanced methods to improve image resolution. Publicly accessible COVID-19 CT image datasets are very difficult to come by due to privacy concerns, which impedes the study and development of AI-powered COVID-19 diagnostic algorithms based on CT images.

**Data description:** To address this problem, we have introduced HRCTCov19, a new COVID-19 high-resolution chest CT scan image dataset that includes not only COVID-19 cases of Ground Glass Opacity (GGO), Crazy Paving, and Air Space Consolidation but also CT images of cases with negative COVID-19. The HRCTCov19 dataset, which includes slice-level, and patient-level labels, has the potential to aid COVID-19 research, especially for diagnosis and differentiation using artificial intelligence algorithms, machine learning, and deep learning methods. This dataset is accessible through the web at: http://databiox.com and includes 181,106 chest HRCT images from 395 patients with four labels: GGO, Crazy Paving, Air Space Consolidation, and Negative.

## Introduction

Coronavirus disease 2019 (COVID-19) is a highly contagious disease that causes severe respiratory distress syndrome. Since 2019, COVID-19 has spread fast throughout several cities in China and other nations (1). According to the WHO, as of February 2022, there have been over 378,000,000 confirmed cases worldwide, with over 5,670,000 fatalities. There are two main methods for determining whether a person is infected with COVID-19 or not: the first uses chest computerized tomography (CT) images, and the second uses a reverse-transcription polymerase chain reaction (RT-PCR) test, which is based on a patient's respiratory samples, such as nasal mucus (2, 3). Although the gold standard for verifying COVID-19 mostly depends on microbiological tests like real-time polymerase chain reaction (RT-PCR) [4], radiological imaging, particularly thin-section CT, is critical. CT is important for the care and follow-up of patients with COVID-19 since it ensures timely treatment and prevents the illness from spreading to other potentially infected persons. It also aids in the assessment of respiratory morbidity (4, 5).

High-resolution CT (HRCT) of the chest is now routinely employed not just in the diagnosis of COVID-19 pneumonia, but also in the monitoring of its course. In the diagnosis of COVID-19 pneumonia, HRCT chest has a sensitivity of 56-98 percent. Ground Glass Opacity (GGO) is the most prevalent finding on chest CT in patients with COVID-19 pneumonia and although is a nonspecific finding, when seen in a typical pattern, is suggestive of COVID-19 pneumonia (6, 7).

Typical imaging findings are peripheral bilateral GGO with or without consolidation or Crazy-Paving pattern, reverse halo signs or other organizing pneumonia (OP)-related findings, and multifocal GGO of the rounded morphology with or without consolidation or visible intralobular lines (Crazy-Paving) (8).

Atypical characteristics include lobar or segmental consolidation in bacterial pneumonia, cavitation in necrotizing pneumonia, and tree-in-bud opacities with centrilobular nodules, which may be seen in a variety of community-acquired infections and aspiration (9).

During the COVID-19 pandemic, radiologists have conducted much research for fast analysis of a huge number of CT images. To overcome these difficulties, many researchers (such as (10, 11)) have established in-depth learning techniques for COVID-19 filtering from CTs. While the findings of this study are quite positive, there are some limitations. Publicly accessible datasets often comprise a small number of individuals and exclude other types of respiratory illnesses to allow for comparison. Additionally, given the present state of medical professionals' involvement in the treatment of COVID-19 patients, it is improbable that they will have time to gather and interpret the COVID-19 CT scan information. Additionally, CT scans might originate from a variety of different sources and be acquired using a variety of different imaging methods, limiting the breadth of an integrated analysis (12-14).

Today, no one is unaware of artificial intelligence's potential in all fields. However, the main difficulty and necessity for the deployment of artificial intelligence-based algorithms is the availability of a vast database of information sufficient to bring artificial intelligence's performance to an acceptable level. Access to huge data sets is problematic in the medical field owing to the high cost of data collecting, patient privacy concerns, and so on. Transfer learning is used to accomplish this. For this purpose, artificial intelligence is taught on a large database and then applied to a smaller database using the trained model. This study's objective is to create a dataset of CT scan images of COVID-19 patients. It is envisaged that academics will utilize the provided dataset in the future to utilize in pre-trained algorithms.

## Data description

The main idea for the organization of this dataset is inspired by a histopathological image dataset for grading breast invasive ductal carcinoma which was conducted and published by DataBioX researchers in 2020 (15). HRCTCov19 dataset consists of 181,106 high-resolution chest images (Grayscale) taken from 395 patients labeled in four categories: GGO (N = 288), Crazy Paving (N = 57), Air Space Consolidation

(N = 27) and Negative (N = 23) amongst 197 Female, 193 Male and 5 Not Specified. Some samples from each category and a summary of this dataset specification are represented in Fig.1 and Table 1, respectively.

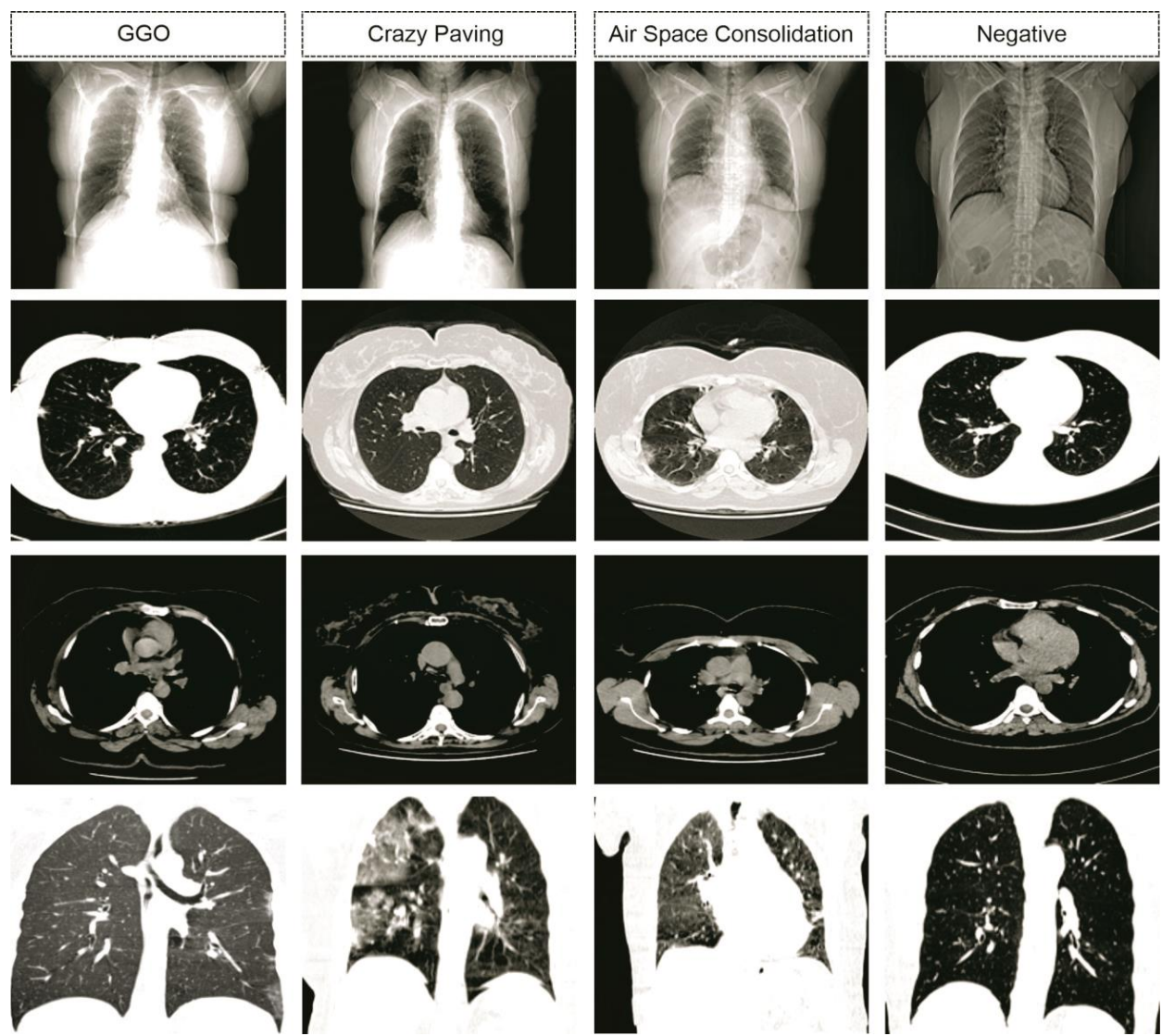

**Fig. 1** HRCTCov19 image dataset samples from each four labels

**Table 1.** A summary of HRCTCov19 dataset specification

| Label | Name of data file/data set | File types (file extension) | Identifier (DOI or accession number) |
|---|---|---|---|
| Data file | 01_HRCTCov19_Dataset.xlsx | Excel file (.xlsx) | Zenodo data (https://doi.org/10.5281/zenodo.10252424)(16) |
| Data set 1 | Air Space Consolidation.rar | RAR file (.rar) | Zenodo data (https://doi.org/10.5281/zenodo.10252424)(16) |
| Data set 2 | Crazy Paving.rar | RAR file (.rar) | Zenodo data (https://doi.org/10.5281/zenodo.10252424)(16) |
| Data set 3 | GGO.rar | RAR file (.rar) | Zenodo data (https://doi.org/10.5281/zenodo.10252424)(16) |
| Data set 4 | Negative.rar | RAR file (.rar) | Zenodo data (https://doi.org/10.5281/zenodo.10252424)(16) |

This dataset was compiled from patients who came to Milad hospitals' emergency rooms between February and September of 2021 year. It includes individuals who received both chest CT and RT-PCR for suspected COVID-19 pneumonia.

All CT examinations were performed with a 128-row Multidetector CT system (SOMATOM Definition Flash, Siemens Healthcare system, Germany). Scanning coverage was from the thoracic inlet to the inferior level of the costophrenic angle. The CT scanning protocol was as follows: tube voltage of 100KV, tube current of 100-200 mA (automatic exposure control employed), rotation time of 0.35 s, pitch of 1.4 mm, detector collimation of 0.6 mm, slice thickness/reconstruction thickness of 5 mm/1 mm. All scans were performed in the supine position during end-inspiration.

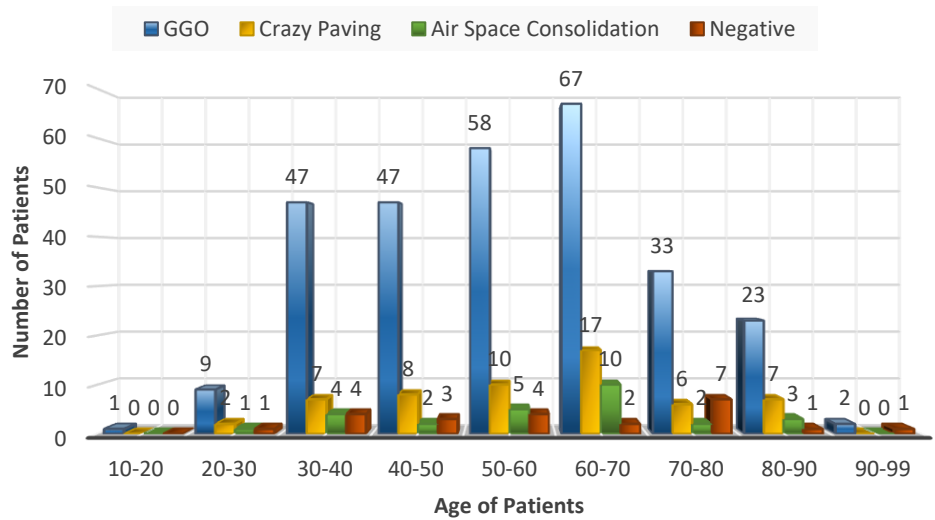

**Fig. 2** Distribution of cases based on their age and infection type in the dataset

Data were transferred to the image processing workstation. These CT images were reviewed on the above-mentioned workstation at the lung window (width of 1000 HU and window level of −700 HU). To ensure the accuracy of the analysis, all images were independently evaluated by an experienced radiologist. The diagnosis of Covid illness was further confirmed by the results of the patients' RT-PCR tests. Figure 2 Shows information about the number of patients with four infections in the dataset.

The COVID dataset presented in other studies has either provided CT data just from COVID patients alone (17), or the data presented falls into two categories: COVID and Normal(13, 18-20).

In a study by Shakuri et al., an open-source repository of more than 1,000 CT scans of COVID-19 lung infection was published and developed by authorized radiologists in the study. From March 2020 to January 2021, CT images were collected from two major university hospitals in Mashhad. Comparative assays, such as RT-PCR and associated clinical signs, were used to confirm COVID-19 infections. The authors state that all data is recorded in the DICOM standard as 16-bit images on a gray scale of $512 \times 512$ pixels (21).

Afshar, et al. introduced the COVID-CT-MD dataset which consists of 169 confirmed positive COVID-19 cases, 76 normal cases, and 60 community-acquired pneumonia (CAP) cases. All these cases were collected from the Babak Imaging Center in Tehran, Iran (12).

Zaffino et al. published a dataset including 50 COVID-19-positive patients in ITK-based file format as a non-contrast chest CT. The presented datasets were obtained from Azienda Ospedaliera Pugliese-Ciaccio in Catanzaro, Italy by two different scanners (22).

In addition, Soares et al. provided a COVID-CT scan dataset from the Public Hospital of the Government Employees of Sao Paulo Metropolitan Hospital of Lapa in Brazil. This dataset includes 210 patients, only 80 of them are infected with COVID-19 and collected from March 15 to June 15, 2020 year. The matrix size of the images was $512 \times 512$ with breath-hold at full inspiration (23).

The database we provide is HRCT images. A CT scan of the chest uses X-rays to obtain images of the lung tissue. The images are obtained in "slices" or thin views that are put together to form an image. The slices of an HRCT are much thinner than with a standard CT scan giving a more detailed image. HRCT scans take one-millimeter slices (24). Given the variety of infection patterns covered in this large dataset, it may be used as a starting point for more extensive data-train-driven AI models.

The limitations we faced during dataset preparation were the missing or defective data, and the poor quality of the images due to breathing artifacts caused by the patient's difficulty breathing. In addition, some categories, including "Crazy Paving" and "Air Space consolidation", contained fewer data compared to the GGO category.

**Abbreviations**

HRCT: High-resolution CT, GGO: Ground Glass Opacity, OP: organizing pneumonia.

**Acknowledgment**
We thank the staff of the imaging department of Milad Hospital, Isfahan, Iran.

**Author contribution**
All of the authors contributed to the design of the work equally, I. A. in data collection, and image classification, M. V. in image classification, and processing works for preparing dataset and writing the manuscript, B. O. in scientific supervision and data collection, M. Z. in data collection, and editing the manuscript, H. B. in processing works, and supervision. All of the authors read, revised, and approved it.
In addition, all authors have agreed both to be personally accountable for the author's contributions and to ensure that questions related to the accuracy or integrity of any part of the work, even ones in which the author was not personally involved, are appropriately investigated, resolved, and the resolution documented in the literature.

**Funding**
No special funding was received for the study.

**Availability of data and materials**
The datasets generated and/or analyzed during the current study are available in the Zenodo repository in (https://doi.org/10.5281/zenodo.10252424). The latest version is also accessible via the DataBioX website: https://databiox.com.

## Declarations

**Ethics approval and consent to participate:** This study has been approved by the ethics committee of the Isfahan University of Medical Sciences. All methods were carried out by relevant guidelines and regulations of the declaration of Helsinki. Informed consent was obtained from all subjects and/or their legal guardian(s).

**Consent for publication:** Not applicable.

**Competing interests:** The authors declare that they have no competing interests.

## References

1. World Health Organization Geneva2020 [Coronavirus disease 2019 (COVID-19) situation report-94:[Available from: https://www.who.int/docs/default-source/coronaviruse/situation-reports/20200423-sitrep-94-covid-19.pdf?sfvrsn=b8304bf0_4.
2. Zhao D, Yao F, Wang L, Zheng L, Gao Y, Ye J, et al. A Comparative Study on the Clinical Features of Coronavirus 2019 (COVID-19) Pneumonia With Other Pneumonias. Clin Infect Dis. 2020;71(15):756-61.
3. Fang Y, Zhang H, Xie J, Lin M, Ying L, Pang P, et al. Sensitivity of Chest CT for COVID-19: Comparison to RT-PCR. Radiology. 2020;296(2):E115-e7.
4. Huang C, Wang Y, Li X, Ren L, Zhao J, Hu Y, et al. Clinical features of patients infected with 2019 novel coronavirus in Wuhan, China. Lancet. 2020;395(10223):497-506.
5. Yang W, Sirajuddin A, Zhang X, Liu G, Teng Z, Zhao S, et al. The role of imaging in 2019 novel coronavirus pneumonia (COVID-19). Eur Radiol. 2020;30(9):4874-82.
6. Chung M, Bernheim A, Mei X, Zhang N, Huang M, Zeng X, et al. CT Imaging Features of 2019 Novel Coronavirus (2019-nCoV). Radiology. 2020;295(1):202-7.
7. Zhou S, Wang Y, Zhu T, Xia L. CT Features of Coronavirus Disease 2019 (COVID-19) Pneumonia in 62 Patients in Wuhan, China. AJR Am J Roentgenol. 2020;214(6):1287-94.
8. Shirani F, Shayganfar A, Hajiahmadi S. COVID-19 pneumonia: a pictorial review of CT findings and differential diagnosis. Egyptian Journal of Radiology and Nuclear Medicine. 2021;52:1-8.
9. Simpson S, Kay FU, Abbara S, Bhalla S, Chung JH, Chung M, et al. Radiological Society of North America expert consensus statement on reporting chest CT findings related to COVID-19. Endorsed by the Society of Thoracic Radiology, the American College of Radiology, and RSNA. Journal of thoracic imaging. 2020.
10. Huang L, Han R, Ai T, Yu P, Kang H, Tao Q, et al. Serial Quantitative Chest CT Assessment of COVID-19: A Deep Learning Approach. Radiol Cardiothorac Imaging. 2020;2(2):e200075.
11. Li L, Qin L, Xu Z, Yin Y, Wang X, Kong B, et al. Artificial intelligence distinguishes COVID-19 from community acquired pneumonia on chest CT. Radiology. 2020.
12. Afshar P, Heidarian S, Enshaei N, Naderkhani F, Rafiee MJ, Oikonomou A, et al. COVID-CT-MD, COVID-19 computed tomography scan dataset applicable in machine learning and deep learning. Sci Data. 2021;8(1):121.
13. Yang X, He X, Zhao J, Zhang Y, Zhang S, Xie P. COVID-CT-dataset: a CT scan dataset about COVID-19. arXiv preprint arXiv:200313865. 2020.
14. Silva P, Luz E, Silva G, Moreira G, Silva R, Lucio D, et al. COVID-19 detection in CT images with deep learning: A voting-based scheme and cross-datasets analysis. Informatics in medicine unlocked. 2020;20:100427.
15. Bolhasani H, Amjadi E, Tabatabaeian M, Jassbi SJ. A histopathological image dataset for grading breast invasive ductal carcinomas. Informatics in Medicine Unlocked. 2020;19:100341.
16. Abedi I, Vali M, Otroshi B, Zamanian M, Bolhasani H. HRCTCov19: A High Resolution Chest CT Scan Image Dataset for COVID-19 Diagnosis and Differentiation. 3 ed. Mendely Data2023.
17. Jacob J, Alexander D, Baillie JK, Berka R, Bertolli O, Blackwood J, et al. Using imaging to combat a pandemic: rationale for developing the UK National COVID-19 Chest Imaging Database. Eur Respir J. 2020;56(2).
18. Morozov SP, Andreychenko A, Pavlov N, Vladzymyrskyy A, Ledikhova N, Gombolevskiy V, et al. Mosmeddata: Chest ct scans with covid-19 related findings dataset. arXiv preprint arXiv:200506465. 2020.
19. Soares E, Angelov P, Biaso S, Froes MH, Abe DK. SARS-CoV-2 CT-scan dataset: A large dataset of real patients CT scans for SARS-CoV-2 identification. MedRxiv. 2020:2020.04. 24.20078584.
20. Rahimzadeh M, Attar A, Sakhaei SM. A fully automated deep learning-based network for detecting COVID-19 from a new and large lung CT scan dataset. Biomedical Signal Processing and Control. 2021;68:102588.

21.	Shakouri S, Bakhshali MA, Layegh P, Kiani B, Masoumi F, Ataei Nakhaei S, et al. COVID19-CT-dataset: an open-access chest CT image repository of 1000+ patients with confirmed COVID-19 diagnosis. BMC Res Notes. 2021;14(1):178.
22.	Zaffino P, Marzullo A, Moccia S, Calimeri F, De Momi E, Bertucci B, et al. An open-source covid-19 ct dataset with automatic lung tissue classification for radiomics. Bioengineering. 2021;8(2):26.
23.	Soares E, Angelov P, Biaso S, Cury M, Abe D. A large multiclass dataset of CT scans for COVID-19 identification. Evolving Systems. 2023:1-6.
24.	Hill LE, Ritchie G, Wightman AJ, Hill AT, Murchison JT. Comparison between conventional interrupted high-resolution CT and volume multidetector CT acquisition in the assessment of bronchiectasis. Br J Radiol. 2010;83(985):67-70.